\documentclass{kluwer}
\usepackage{graphicx,psfig}
\begin{document}
\begin{article}
\begin{opening}
\title{Chemodynamical evolution of interacting galaxies}
\author{F. \surname{Combes} and A.L. \surname{Melchior}}
\institute{Observatoire de Paris, DEMIRM,
61 Av. de l'Observatoire, F-75014, Paris, FRANCE}
\vspace{-0.5cm}
\begin{abstract}
We have undertaken numerical simulations of galaxy interactions 
and mergers, coupling the dynamics with the star formation history
and the chemical evolution. The self-gravity of stars and gas is
taken into account through a tree-code algorithm, the gas
hydrodynamics through SPH, and an empirical law such as
a local Schmidt law is used to compute star formation. The gas 
and stellar metallicity is computed at each position, according to 
assumed yields, and the dust amount is monitored. At each step
the spectra of galaxies are computed, according to simple 
radiative transfer and dust models. Initial conditions for these
simulations will be taken from a large-scale cosmological frame-work.
The aim is to build a statistically significant library of merger
histories. The first results of the project will be discussed, in 
particular on predictions about galaxy surveys at high redshift. 
\end{abstract}
%\keywords{Galaxies: evolution}
%
\end{opening}
\vspace{-0.8cm}
\section{Introduction}
\vspace{-0.4cm}
Although the quiescent or ``steady-state'' mode of star formation is dominant
today at $z=0$, it appears that 
most of the star formation in the Universe might have occurred in starbursts.
In this violent second mode of star formation, the 
rate can be one or two orders of magnitude larger than in the quiescent mode,
during short periods of time of $\sim 10^8$ yr.
The fraction of starbursts increases with redshift, as revealed
for instance by the submillimetric surveys (e.g. Carilli et al 2000)
and  the star formation rate increases as $(1+z)^m$,
where $m$ is a power between 3 and 4, as shown by the Madau
curve (see the review by Genzel \& Cesarsky 2000).
This is due first to the frequency of mergers triggering starbursts,
 that increases with a high power (Le F\`evre et al. 2000);
and second to the much larger efficiency of mergers to form stars, which 
can be explained by: \\
-- 1-  a larger gas fraction in young galaxies 
-- 2- smaller dynamical time-scales, since the first haloes to form are the densest 
-- 3- more unstable galaxies, since the bulge-to-disc ratio increases with time.

To determine which are the most important parameters in the z-evolution
of mergers, we have undertaken numerical simulations. Our aim is to 
couple dynamics with star formation, chemical enrichment, 
multiphase physics of the interstellar medium, and to deduce
the observational properties of these mergers, computing their
total spectral energy distribution (SED) from stellar, gas and dust emission.

\begin{figure}
\psfig{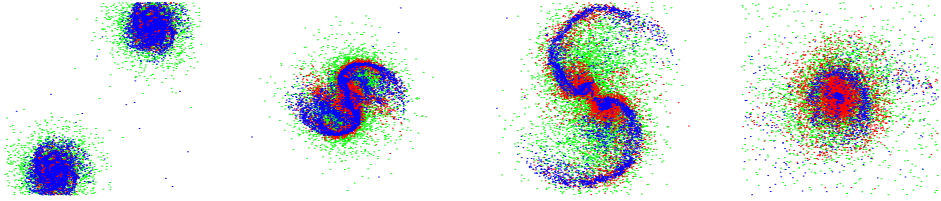}
\caption{Example of a simulation run, involving two equal mass spiral galaxies,
in a close-by encounter, leading to a merger. Dark matter particles are in green,
gas in blue and stars in red (cf http://wwwusr.obspm.fr/$\sim$demirm/fc-fig1.gif)}
\end{figure}

\vspace{-0.8cm}
\section{The numerical model} 
\vspace{-0.4cm}
The gravitational dynamics of the interactions is simulated with 
a TREE-SPH code;
since the aim is to gain a statistical view with a large number
of parameters explored, the number of particles is modest,
24 000, divided into 
stars : 8000, gas : 8000,  and dark matter : 8000.
The preliminary results presented here do not include
the collisional molecular clouds (simulated by a sticky particles
code).
The smallest timescale is of the order of  1 Myr, the
resolved sizes between 150 pc and  0.5 kpc, and the  mass scale 
of 10$^6$ M$_\odot$ for the gas (column gas density ranges
from 2 10$^{20}$ to 10$^{23}$ cm$^{-2}$).

The adopted star formation recipe is a ``local'' Schmidt law (i.e.
at the resolution scale): the
fraction of any gas particle transformed to stars
is $\propto \rho^{n-1}$, where
$\rho$ is the volumic gas density, and $n$ the power of the Schmidt law.
To avoid a variable number of particles, 
the algorithm of hybrid particules is used
(cf Mihos \& Hernquist 1994): some particles are
transiently partly gaseous, and partly stellar, until their gas fraction
drops below 5\%; they are then
turned into pure stars, the gas being spread among the neighbours.

Two possibilities have been explored for the stellar mass loss:

-- First, as often done, instantaneous recycling is assumed; most of
the gas is re-injected by the massive stars with a short life-time, in 
the first few Myrs. With a
Scalo IMF, with a mass spectrum from 0.1 to 100 M$_\odot$, 
9\% of the gas turned into stars is re-injected in the ISM
quasi- instantaneously.
The adopted yield (the ratio of the mass of ejected metals to the stellar mass)
is $y  = 0.02$.  The
energy of supernovae and stellar winds is partly reinjected in the ISM
under the form of kinetic energy,  
through expanding velocity of the surrounding gas.

-- Second, the mass loss is now spread over much larger time-scale
(Gyrs), and up to 40\% of the stellar mass can be lost, with a loss-rate
approximated by an $\propto 1/t$ law (Jungwiert, Combes \& Palous 2001).
This brings significant changes, since large amounts of gas become
available where it was consumed out in a burst.

\begin{figure}[h]
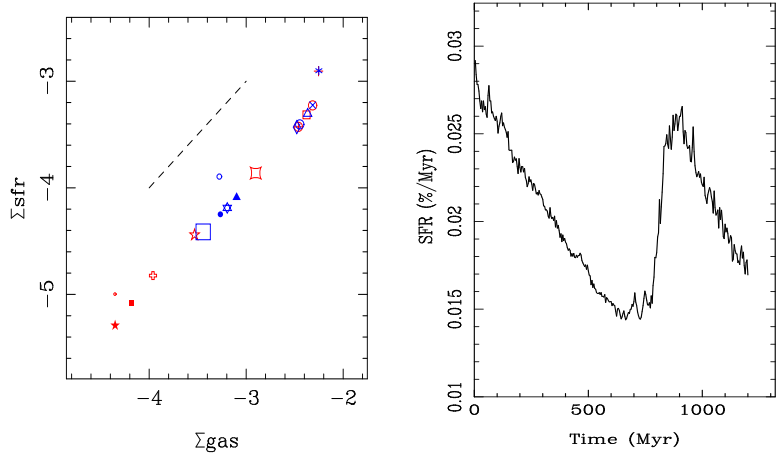

\begin{tabular}{cc}
\psfig{width=5cm,height=6cm,file=combesf.fig2a.ps,bbllx=5.5cm,bblly=1cm,bburx=17.5cm,bbury=12cm,angle=-90} &
\psfig{width=5cm,height=6cm,file=combesf.fig2b.ps,bbllx=3cm,bblly=1cm,bburx=20.5cm,bbury=12cm,angle=-90}\\
\end{tabular}
\caption{{\bf Left:} Derived ``global'' Schmidt law from a merger simulation run, where 
the various points correspond to the various epochs of a simulation (where $n=1.2$), for
the two galaxies separately (red and blue). 
The slope of $n=1$ is indicated by the dash line.
{\bf Right:} Star formation rate as a function of time for the same
run. The second peak follows the closest approach leading to merging, at t=700 Myr}
\end{figure}

\vspace{-0.8cm}
\section{First results}
\vspace{-0.4cm}
A simulation of two spiral discs
merging at late redshift is plotted in fig 1.
We first test the star formation recipe: does it reproduce 
the "global" (i.e. over the whole galaxy) Schmidt law observed 
when the total surface density of stars formed
over the disc is compared to the average gas surface density
 (Kennicutt 1998)? To check that, the radii
of the galaxies were computed as those containing 90\% of the baryonic 
mass, and only the recent stars (formed over the last 100 Myr) were
considered. A linear law is observed in most cases (fig 2, left), whatever
the power of the ``local'' Schmidt law, provided that $n\le 1.5$. 
For $n=2$, this was not verified.

The star formation rate as a function of time
reveals one or two peaks of starburst activity,
depending on the relative orbits between the merging galaxies
(fig 2, right). Starbursts are concentrated in the nuclei.
Gradients of metallicity are washed out in the discs 
by the merger, although there still is a high metallicity peak
in the nuclei.

\vspace{-0.8cm}
\section{Observations of the mergers}
\vspace{-0.4cm}

The results of the simulation are "observed" at different
wavelengths, from the millimeter to optical, with a "pixel" size
adapted to present or future instruments (NGST, ALMA, etc..).
To obtain the SED of the galaxy systems, at each epoch
and for each pixel, we compute a synthesis of stellar population
according to their formation age and metallicity 
(from PEGASE, Fioc and Rocca 1997).
This distribution of populations  gives the unobscured optical SED.
Then, from the gas in front of stars, its column density and metallicity,
the extinction is computed for each pixel, and
all the luminosity absorbed is re-radiated by dust in the far-infrared
domain (cf fig 3). The dust model is derived from
that of the Milky Way, and local starbursts
(Melchior et al. 2001).

\vspace{-0.8cm}
\section{Preliminary conclusions}
\vspace{-0.4cm}

Although the  limited spatial resolution prevents to
take into account the details of star formation, the large-scale dynamical
triggering mechanisms appear to be taken into account rather realistically.
Up to now, only isolated mergers have been considered, which 
suffer from gas fueling problems. In the future,
the systems will accrete external gas, depending on environment.
In the next step, each simulation will be coupled to larger-scale 
cosmological simulations, providing boundary conditions.

%\resizebox{0.62\textwidth}{!}
\begin{tabular}{cc}
   {\psfig{file=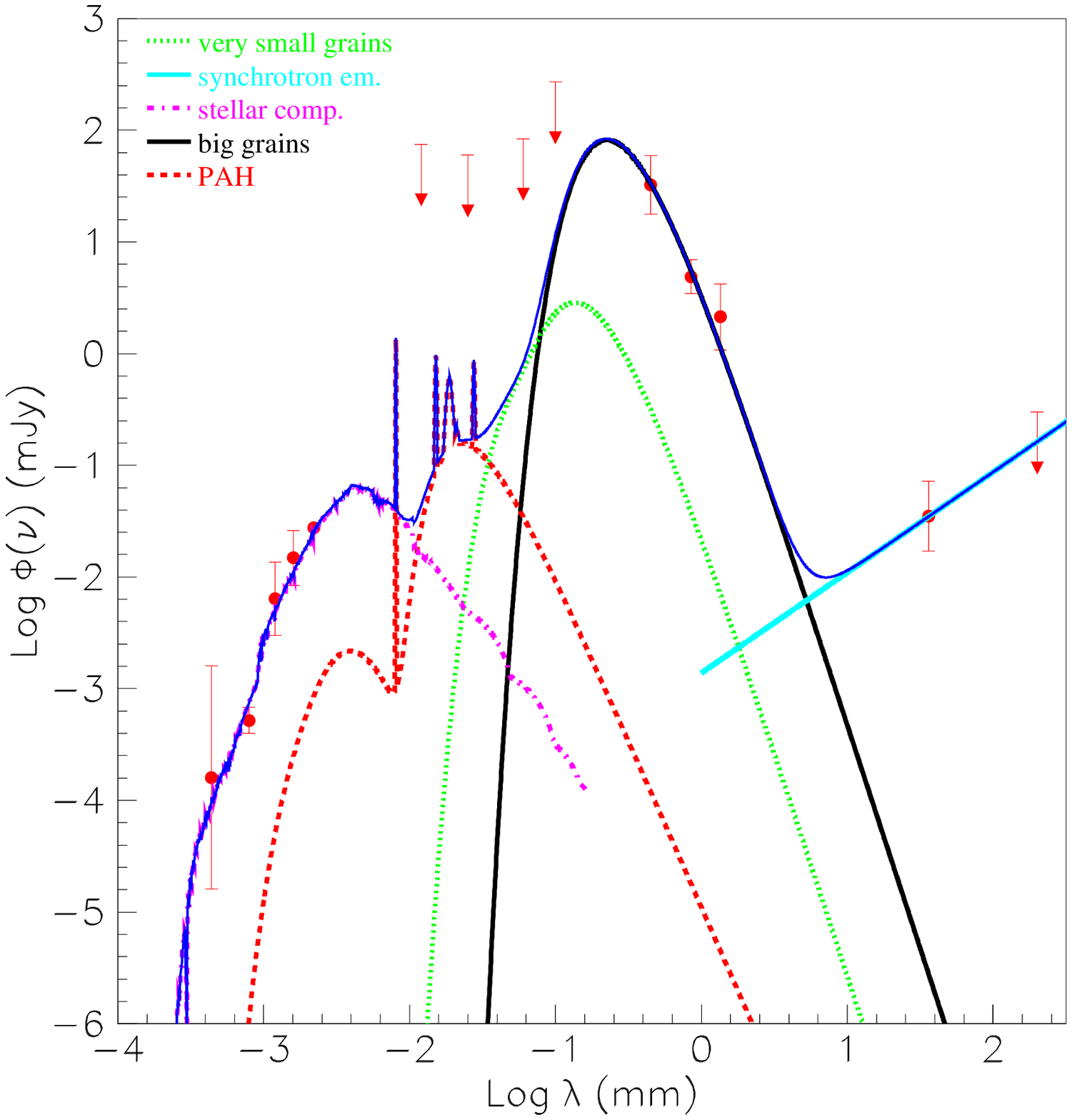,width=0.72\textwidth,clip=}} &
% \hfill
\parbox[b]{0.38\textwidth}{\scriptsize
%\caption{
{\it Figure 3.} Application of our SED modelling
to HR10 (Dey et al. 1999). We 
developed a Simple Stellar Population library derived
from the PEGASE package (Fioc \& Rocca-Volmerange 1997) for different
ages and metallicities. For each element of resolution (e.g. pixel),
we compute a spectrum as follows. With the stellar age and metallicity,
we choose the stellar emission from this library. We add an empirical
extinction -- constrained by nearby starbursts by Calzetti et
al. (2000) -- on this stellar flux. The absorbed flux is reemitted in
the IR. Relying on D\'esert et al. (1990), we distribute this flux
into 3 dust components: big grains, small grains and PAH.\\
\vspace{1cm}\\
}\\
\end{tabular}
%}

\end{article}
\end{document}